\documentclass[aip, apl,reprint,]{revtex4-1}
\usepackage{graphicx}
\usepackage{bm}
\usepackage{dcolumn}
\usepackage[hidelinks]{hyperref}
\usepackage{gensymb}
\usepackage{dcolumn}    
\usepackage{ifpdf}
\usepackage{amssymb,lineno,amsfonts}
\usepackage{graphicx}   
\usepackage{bm}         
\usepackage{bbm}
\usepackage{mathrsfs}
\usepackage{upgreek}
\usepackage{epstopdf}
\usepackage{setspace}
\usepackage{hyperref}
\usepackage{esvect}

\usepackage[usenames,dvipsnames]{xcolor}
\definecolor{med-blue}{RGB}{25,25,112}
\hypersetup{colorlinks, linkcolor={blue},citecolor={blue}, urlcolor={blue}}
\usepackage{times}
\usepackage [english]{babel}
\usepackage [autostyle, english = american]{csquotes}\usepackage{graphicx}
\usepackage{bm}
\usepackage{dcolumn}
\usepackage[hidelinks]{hyperref}
\usepackage{gensymb}
\usepackage{dcolumn}    
\usepackage{ifpdf}
\usepackage{amssymb,lineno,amsfonts}
\usepackage{graphicx}   
\usepackage{bm}         
\usepackage{bbm}
\usepackage{mathrsfs}
\usepackage{upgreek}
\usepackage{epstopdf}
\usepackage{setspace}
\usepackage{hyperref}
\usepackage{esvect}

\usepackage[usenames,dvipsnames]{xcolor}
\definecolor{med-blue}{RGB}{25,25,112}
\hypersetup{colorlinks, linkcolor={blue},citecolor={blue}, urlcolor={blue}}

\begin{document}

\preprint{AIP/123-QED}

\title  {\textcolor{blue}{Magnetic and dielectric investigations of $\gamma$ - Fe${_2}$WO${_6}$}}

\author{Soumendra Nath Panja}
\author{Jitender Kumar}
 \author{Luminita Harnagea}
\affiliation{Department of Physics, Indian Institute of Science Education and Research,\\ Dr. Homi Bhabha Road, Pune, Maharashtra-411008, India}%
\author{A. K. Nigam}
\affiliation{Department of Condensed Matter Physics and Material Science, Tata Institute of Fundamental Research, Dr. Homi Bhabha Road, Mumbai 400 005, India}
\author{Sunil Nair}
\email{Author to whom correspondence should be addressed. Electronic mail:
	sunil@iiserpune.ac.inr}
\affiliation{Department of Physics, Indian Institute of Science Education and Research,\\ Dr. Homi Bhabha Road, Pune, Maharashtra-411008, India}
\affiliation{Centre for Energy Science, Indian Institute of Science Education and Research,\\ Dr. Homi Bhabha Road, Pune, Maharashtra-411008, India}
\date{\today}

\begin{abstract}
The magnetic, thermodynamic and dielectric properties of the $\gamma$ - Fe${_2}$WO${_6}$ system is reported. Crystallizing in the centrosymmetric $Pbcn$ space group, this particular polymorph exhibits a number of different magnetic transitions, all of which are seen to exhibit a finite magneto-dielectric coupling. At the lowest measured temperatures, the magnetic ground state appears to be glass-like, as evidenced by the waiting time dependence of the magnetic relaxation. Also reflected in the frequency dependent dielectric measurements, these signatures possibly arise as a consequence of the oxygen non-stoichiometry, which promotes an inhomogeneous magnetic and electronic ground state.   
\end{abstract}

\maketitle
The area of magneto-dielectrics - which pertains to the coupling between the magnetic and dielectric properties - has seen a renaissance in the recent past. This is partly due to emergence of the area of magnetoelectric multiferroics, where magnetic and polar orders co-exist. In these systems, the onset of ferroelectric order typically results in a pronounced dielectric anomaly, which can then be tuned by the application of an external magnetic field\cite{NHur,Eerenstein,khomskii} . However, the phenomena of magnetodielectricity is more generic, since it is not constrained by the stringent symmetry considerations which are a prerequisite for the observation of either magnetoelectricity or multiferroicity.  A number of potential applications varying from spin-charge transducers to magnetic sensors can be envisaged using magnetodielectric materials \cite{Yang,GSrinivasan} and this area of research is continuously driven by the investigation of different material and structural classes which could exhibit these properties, especially near room temperatures.
Strongly correlated magnetic oxides offer a natural playground for the investigation of such phenomena,  since many of them exhibit an insulating (or at-least a semiconducting) antiferromagnetic ground state. Moreover, the large coupling between the spin, charge and lattice degrees of freedom observed in many of these systems is an added advantage, and typically contributes towards a larger magneto-dielectric effect \cite{LawsMD,RamSeshadri,RseshadriMn3O4,RMahendiran,Palstra}. With the dielectric constant being susceptible to changes in the magnetic structure, it is not surprising that a number of magnetic oxides exhibit magneto-dielectricity in the vicinity of their magnetic transitions. Here we report on the magnetic, thermodynamic and dielectric investigation of a relatively unexplored Iron-Tungsten-Oxygen system (Fe${_2}$WO${_6}$). In addition to a complex set of magnetic transitions, including a low temperature glass like magnetic state, we also observe the existence of a finite magneto-dielectric coupling, persisting right up to room temperatures.   
   
The chemical phase diagram of the Fe-W-O system is characterized by the presence of a number of polymorphic modifications, which makes the selective synthesis of Fe${_2}$WO${_6}$ non-trivial \cite{Leiva} .Prior structural investigations have revealed that Fe${_2}$WO${_6}$ can exist in three distinct structures, depending on their synthesis conditions \cite{Walczak1992,Thomas,Sieber} . Labeled as $\alpha$, $\beta$ and $\gamma$- Fe${_2}$WO${_6}$, these polymorphs are typically stabilized as a function of increasing reaction temperatures, with ill-defined phase boundaries. For instance, $\alpha$-Fe${_2}$WO${_6}$ crystallizing in the orthorhombic columbite ($Pbcn$) symmetry is stabilized at reaction temperatures lower than 800$\degree$C. At reaction temperatures between 750-900$\degree$C, the monoclinically distorted $\beta$-Fe${_2}$WO${_6}$ is favored, whereas at reaction temperatures in excess of 900$\degree$C, the $\gamma$ phase is reported to be stabilized. This high temperature $\gamma$ phase is known to crystallize in the tri-$\alpha$-PbO${_2}$ structure, where the orthorhombic $Pbcn$ symmetry of the $\alpha$- Fe${_2}$WO${_6}$ phase is preserved, but with a tripling of the unit cell along one of the crystallographic directions ($a'=a, b'=3b, c'=c$). 

Polycrystalline specimens of $\gamma$- Fe${_2}$WO${_6}$ were synthesized using the standard solid state ceramic method. An equimolar mixture of previously preheated Fe${_2}$O${_3}$ (Sigma Aldrich, $ \geq$ 99$\% $) and WO${_3}$ (Alfa Aesar, $ \geq$ 99.8$\% $) precursors, were thoroughly ground for several hours using a dry ball mill. The fine and homogenous mixture was pressed into pellets and loaded into a preheated alumina boat. The charge was slowly heated to 800$\degree$C, kept there for 24 hours and then gradually cooled down to room temperature. These pellets were repeatedly reground, pelletized and sintered for several times at 950$\degree$C in air. After about 100 hours at this temperature we obtained a well crystallized single phase of $\gamma$- Fe${_2}$WO${_6}$. Phase purity was confirmed using X-Ray powder diffraction measured using a Bruker D8 Advance diffractometer with Cu K$_{\alpha}$ source, and Rietveld refinement was carried out using the Fullproof refinement program\cite{Fullprof}. These $\gamma$- Fe${_2}$WO${_6}$ specimens were observed to slightly degrade with time, though no appreciable changes were observed in their XRD patterns taken after a few months. Specific heat and magnetization measurements were performed using a Quantum Design PPMS and a MPMS-XL SQUID magnetometer respectively. Temperature dependent dielectric measurements were performed in the standard parallel plate geometry, using a NOVOCONTROL (Alpha-A) High Performance Frequency Analyzer. Measurements were typically done using an excitation ac signal of 1V at frequencies varying from 100 Hz to 100 kHz. Magneto-dielectric measurements were performed by using the Manual Insertion Utility Probe of the MPMS-XL magnetometer. 

\begin{figure}
	\centering
	\hspace{-0.6cm}
	\includegraphics[scale=0.27]{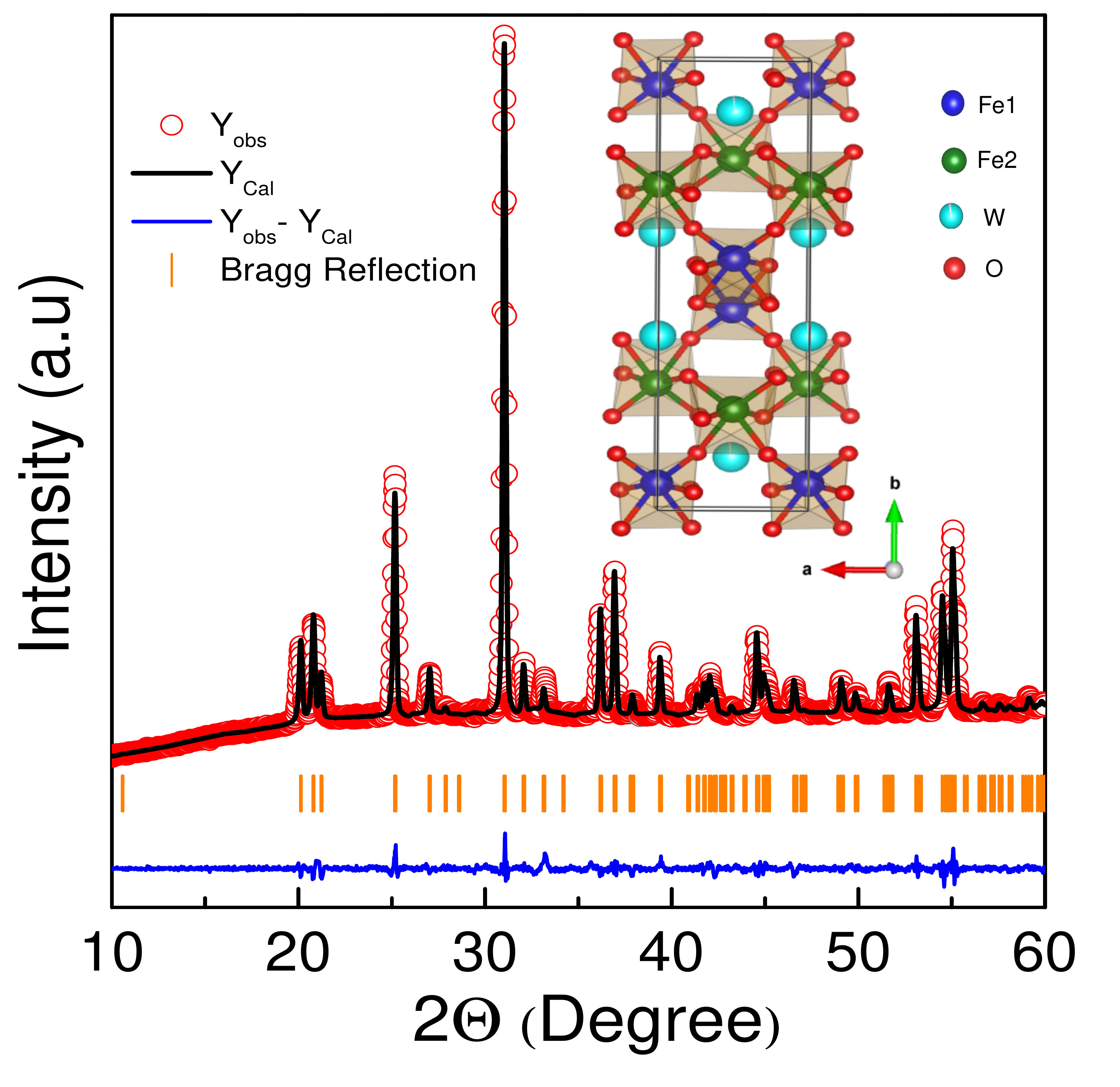}
	\caption{A Rietveld fit to the room temperature x-ray diffraction data of $\gamma$-Fe${_2}$WO${_6}$. This corresponds to a fit with $R$ parameters of $R{_{wp}}$ = 11.3, $R{_e}$ = 6.10 . The crystal structure of this system as viewed along the crystallographic $c$ axis is shown in the inset.}
	\label{Fig1}
\end{figure}

The Rietveld refinement of room temperature X-ray diffraction pattern of our specimen is as shown in Fig.(\ref{Fig1}). 
A good fit, corresponding to goodness of fit value ($R_wp/R_e$ ) of 1.85 could be obtained, confirming a single phase $\gamma$- Fe${_2}$WO${_6}$ crystallizing in the $Pbcn$ space group, with lattice parameters $a = 4.575(1){\AA}$, $b = 16.747(4){\AA}$, $c = 4.965(1){\AA}$, and $\alpha=\beta=\gamma=90\degree$. A schematic of this tri-$\alpha$-PbO${_2}$ columbite structure is depicted in the inset of Fig.(\ref{Fig1}), and comprises of layered corner-shared FeO${_6}$ octahedra, with Fe occupying two distinct crystallographic sites. 

\begin{figure}
	\centering
	\includegraphics[scale=0.30]{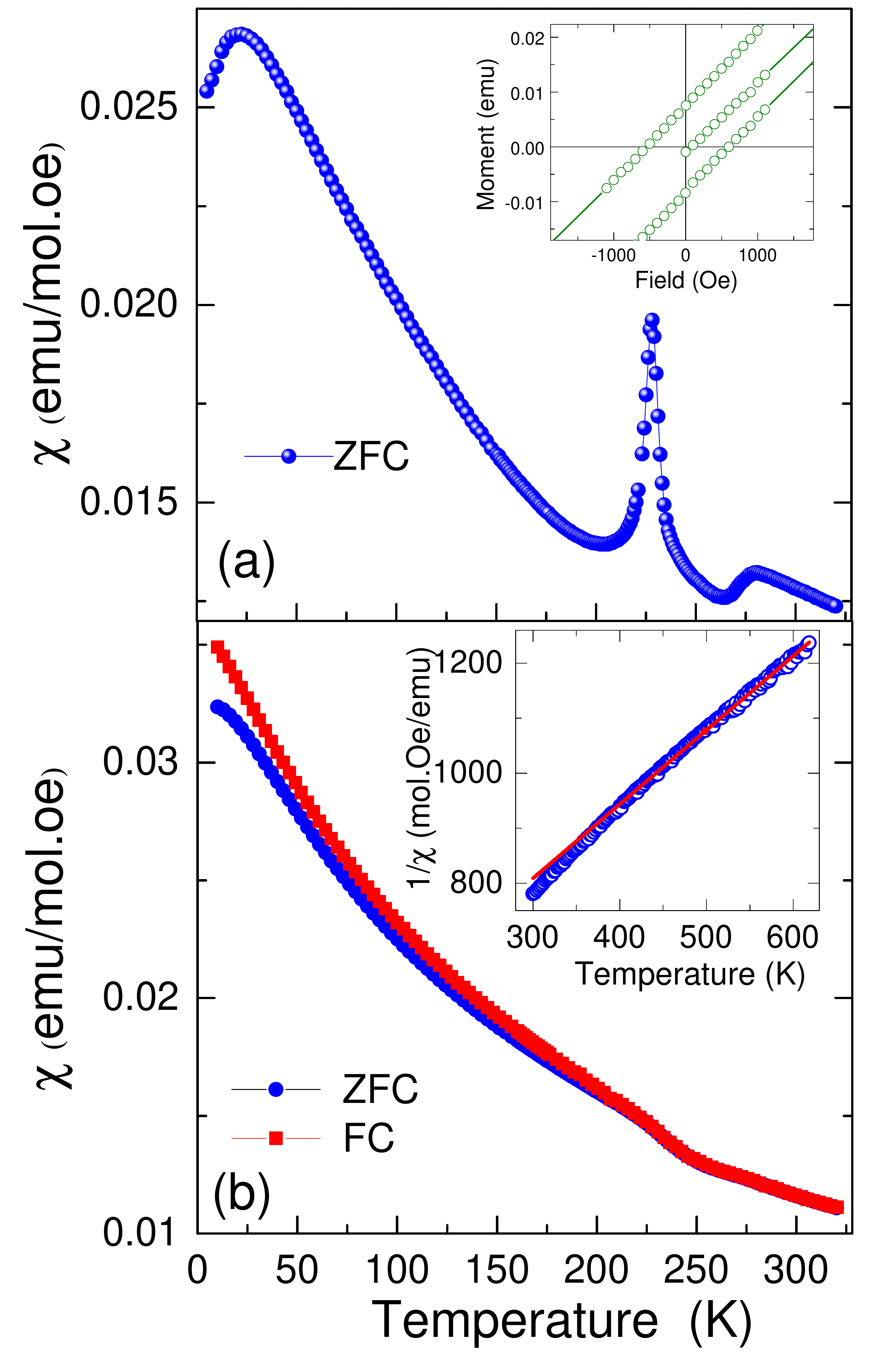}
	\caption{(a) depicts the temperature dependence of dc magnetization as measured in $\gamma$-Fe${_2}$WO${_6}$ in the Zero Field Cooled (ZFC) protocol at an applied field of 10 Oe. The inset depicts an expanded view of the $M$-$H$ isotherm at 2K. (b) depicts the ZFC and FC measured at an applied field of 1 Tesla. The inset shows the Curie-Weiss fir to the high temperature magnetization data}
	\label{Fig2}
\end{figure}

Preliminary magnetic characterization of the Fe${_2}$WO${_6}$ polymorphs have been reported earlier, and appears to depend acutely on the synthesis conditions \cite{GuskosJSSC,MagStruc}. For instance, both the $\alpha$ and $\gamma$-Fe${_2}$WO${_6}$ are reported to exhibit two magnetic transitions, with a high temperature transition at $T{_1}$ $\approx$ 240-260K, and a lower temperature transition $T{_2}$ $\approx$ 200-220K. In addition, the presence of an additional low temperature feature at $\approx$ 20K has also been reported in both these polymorphs. However, the $\beta$-Fe${_2}$WO${_6}$ is reported to exhibit a solitary magnetic transition at 260K. The dc magnetic susceptibility of our  $\gamma$-Fe${_2}$WO${_6}$ specimen, as measured in the Zero Field Cooled (ZFC) and Field Cooled (FC) measuring protocols is shown inset of Fig.\ref{Fig2}(a). 
As is evident from the main panel of \ref{Fig2}(a), two distinct transitions at 280K, and 228K can be discerned from the magnetization measurement performed at low magnetic fields of the order of 10 Oe.  An increase in the applied magnetic fields appears to broaden the high temperature transitions \ref{Fig2}(b), without a pronounced change in the transition temperatures.  An early neutron diffraction investigation of $\gamma$-Fe${_2}$WO${_6}$ has suggested that the magnetic structure comprises of ferromagnetic (100) planes coupled antiferromagnetically, with the spins lying along the (001) direction\cite{MagStruc} . It was also speculated that the magnetic space group ($Pbc'n'$) could allow for a finite ferromagnetic component along one of the crystallographic axes. Magnetic field isotherms measured in our specimen at different temperatures indicate that though the magnetization does not saturate up to the highest measured field of 7 Tesla, a finite opening of the loop exists all the way down to the lowest measured temperatures (inset of \ref{Fig2}(a)). This suggests that a weak ferromagnetic component co-exists along with the antiferromagnetic order which appears to set in at 280K. On further reducing the temperature, the magnetization increases continuously, and exhibits a cusp like feature at $\approx$ 22K, which possibly corresponds to the low temperature transition that some earlier reports have alluded to. Interestingly, even up to magnetic fields up to 1 Tesla \ref{Fig2}(b), a bifurcation in the FC and ZFC magnetization is observed at very low temperatures, indicating that this low temperature feature could possibly be associated with a magnetic glass like state. A Curie Weiss fit to the high temperature DC magnetization data (inset of \ref{Fig2}(b)) at temperatures in excess of 450 K, giving an effective magnetic moment per Fe ion of 1.22$\mu{_{\beta}}$. The spin only moment of (low spin) Fe${^{3+}}$ ion is 1.73$\mu{_{\beta}}$, implying that a finite fraction of the Fe ions are in the low spin ($S =0$) Fe${^{2+}}$ state. An early Electron Paramagnetic Resonance (EPR) measurement has also speculated on the possibility of the existence of  Fe${^{2+}}$ ions in the Fe${_2}$WO${_6}$ system \cite{GuskosJSSC}, indicating that oxygen non-stoichiometry could be endemic to this class of materials. 

This low temperature state was further evaluated by means of magnetic relaxation measurements. These measurements were performed by cooling the system from room temperatures (in the zero field cooled protocol) down to the lowest temperature of 2 K, followed by soaking the system at different waiting times (100, 1000 and 5000 seconds). The variation in the DC magnetization as a function of time was then recorded, on the application of a magnetic field of 100 Oe. 
\begin{figure}
	\centering
	\vspace{-0.5cm}
	\includegraphics[scale=0.45]{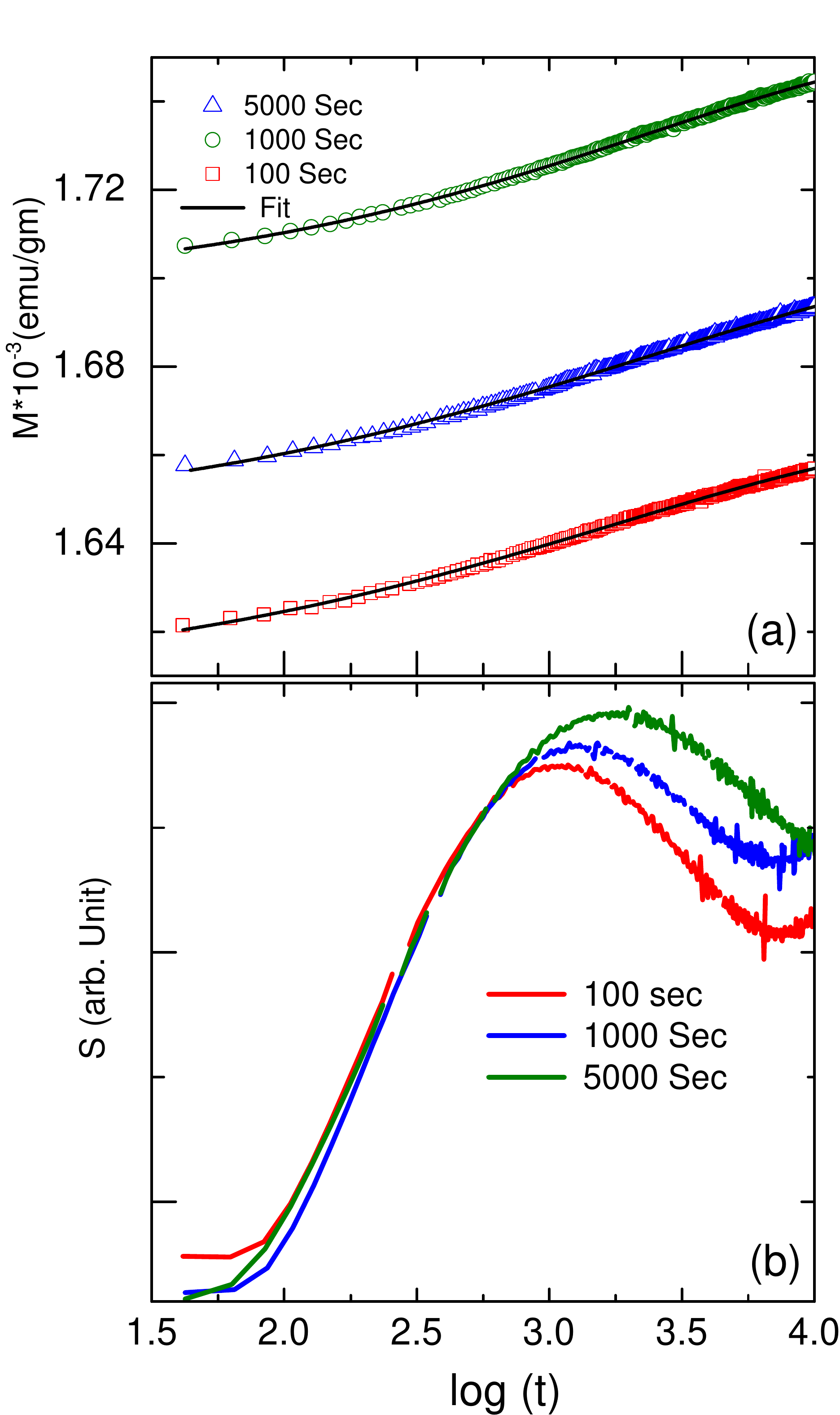}
	\caption{(a) shows the evolution of the dc magnetisation at an applied field of 100 Oe, after soaking the system for different waiting times. The solid lines are fits to a stretched exponential function. (b) depicts the shift of the magnetic viscosity $S(t)$ as a function of the waiting time. }
	\label{Fig3}
\end{figure}
Fig.\ref{Fig3}(a) shows the evolution of magnetization as a function of time in  $\gamma$- Fe${_2}$WO${_6}$. We observe that the data fits well to a stretched exponential of the form $M(t) = M{_0}-M{_g}{exp(\frac{-t}{\tau}})^{\beta}$ which has been used earlier in fitting the magnetic relaxation in a number of spin/cluster glasses \cite{Kohlra,Williams,FreitasStrechExpo}  . Here, $\beta$ is a temperature dependent exponent, $\tau$ is the time constant, and $M{_0}$ and $M{_g}$ refer to the contributions from the long range ordered and glassy components of the system respectively. These fits are depicted as solid lines in Fig.\ref{Fig3}(a), and the values of $M{_0}$ and $\beta$ deduced from our fit  range from 0.00167 to 0.00175 emu/gm and 0.35 to 0.41 respectively for different waiting times. The small value of $M{_0}$ in our case could be a consequence of the fact that the long range ordered component in our case is antiferromagnetic\cite{MoHighvalue}. The magnetic viscosity ($S (t)=1/H(\partial M/\partial (lnt)$))  \cite{Lundgren,SMajumdar,SGiriEPL} as deduced for measurements with different waiting times is depicted in Fig.\ref{Fig3}(b). As is expected for glassy systems, a peak in $S(t)$ which shifts as a function of the waiting time is observed, reinforcing the glass like nature of this low temperature state. As mentioned earlier, the  $\gamma$- Fe${_2}$WO${_6}$ specimens appear to slightly change with time. Our measurements indicate that even after a few months after synthesis, the temperatures of the high temperature magnetic transitions remain relatively invariant. However, the temperature of the low temperature glass like transition appears to shift to lower temperatures as a function of the elapsed time. 

\begin{figure}
	\centering
	\includegraphics[scale=0.40]{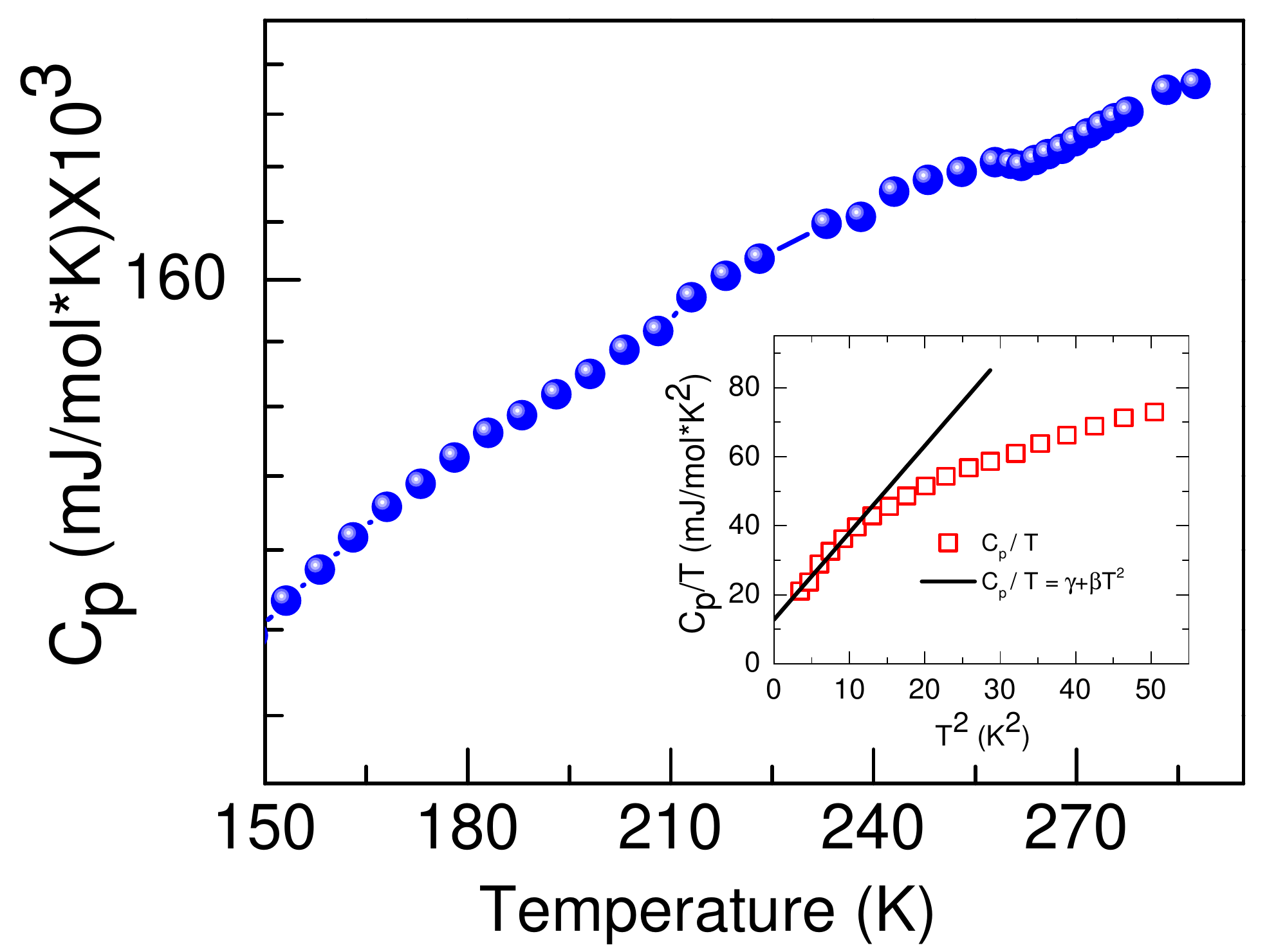}
	\caption{The specific heat of $\gamma$-Fe${_2}$WO${_6}$ is shown in the main panel. The inset depicts $C/T$ as a function of $T{^2}$, with the low temperature fit indicating that at the lowest measured temperatures, the data can be described by the equation $C/T$ = $\gamma$+$\beta T{^2}$. The values of $\gamma$ and $\beta$ obtained from this fit are 12.7 $\pm0.6$ mJ/mol K${^2}$ and 2.52 $\pm0.07$ mJ/mol K${^2}$ respectively.}
	\label{Fig4}
\end{figure}

Specific heat measurements on $\gamma$-Fe${_2}$WO${_6}$ reveals a solitary feature in the vicinity of the high temperature transition, as is shown in the main panel of Fig.(\ref{Fig4}). The fact that the other two transitions are not picked up in the heat capacity measurements indicate that the change in entropy associated with these transitions is quite small. The inset of Fig.(\ref{Fig4}) shows the linear fit to the low temperature specific heat using the equation $C/T = \gamma + \beta T{^2}$, giving values of 12.7 $\pm0.6$ mJ/mol K${^2}$ and 2.52 $\pm0.07$ mJ/mol K${^2}$  for $\gamma$ and $\beta$ respectively. The value of the Debye temperature ($\theta{_D}$) using the relation $\theta{_D}{^3}$ = 234$R/\beta$ (with $R$ being the universal gas constant) gives a value of 91.7 K.

\begin{figure}
	\centering
	\includegraphics[scale=0.40]{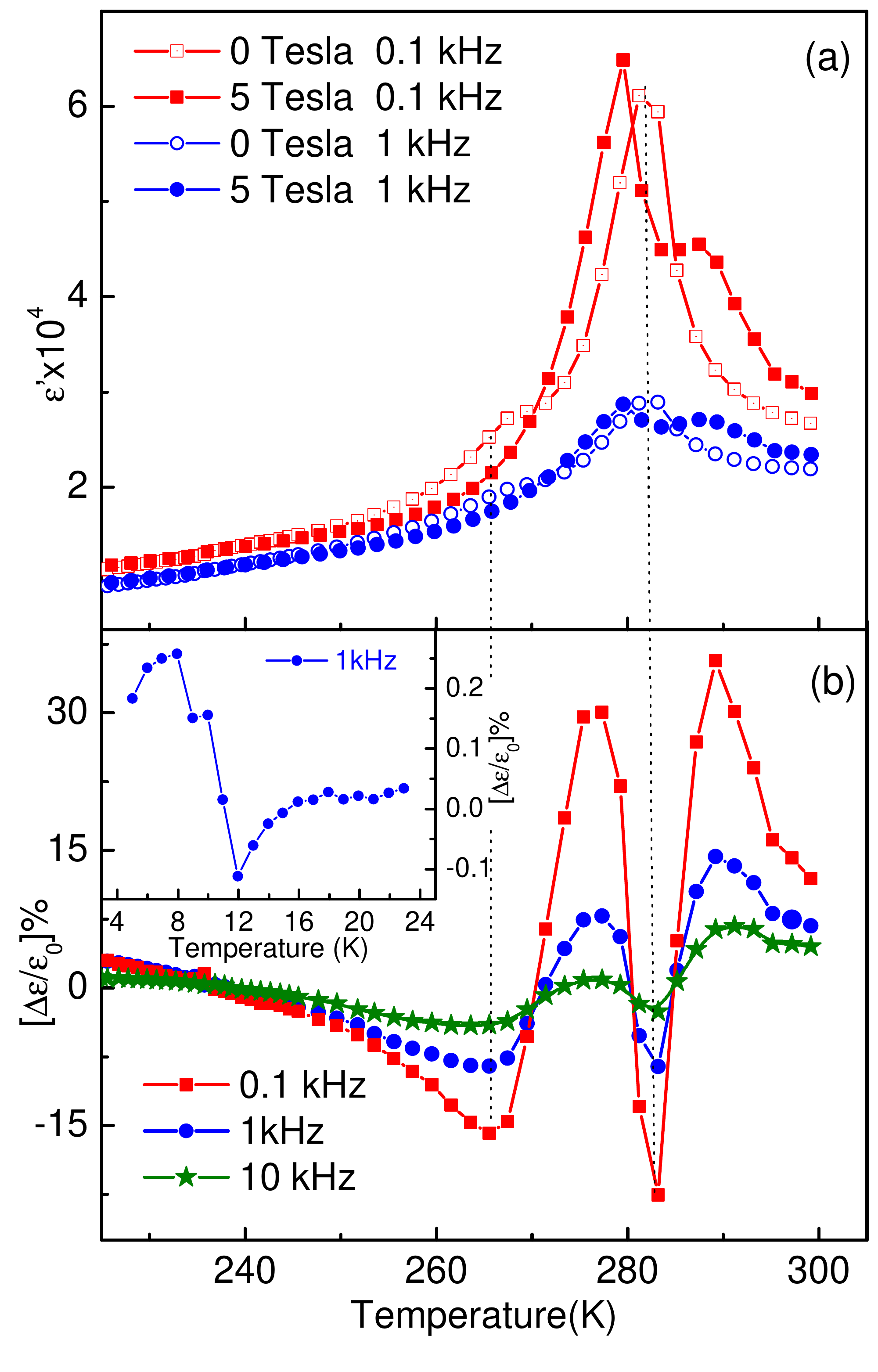}
	\caption{(a) The magnetic field dependence of the real part of the dielectric susceptibility $\epsilon'(T)$ as measured at 0 and 5 Tesla magnetic field. (b) depicts the data plotted as the extent of magneto-dielectricity $|{\Delta \epsilon}/\epsilon{_0}| \%$. The inset shows the same quantity in the vicinity of the low temperature glass transition.}
	\label{Fig5}
\end{figure}

To the best of our knowledge, there have been no reports of dielectric measurements of any of the Fe${_2}$WO${_6}$ polymorphs. Fig.(\ref{Fig5}) depicts the real part of the dielectric constant as a function of temperature, as measured at frequencies varying from 0.1 to 10 kHz. A pronounced dielectric anomaly is observed in the vicinity of the the high temperature magnetic transition at 280K, indicating a substantial coupling between the spin and electronic degrees of freedom in this system. Though the observed dielectric anomaly is reminiscent of that seen in many multiferroic systems, the overall resistivity values are quite low, especially in the high temperature regime, where these features are observed. Thus, the possibility of a magnetically induced ferroelectric state can be ruled out. In this context, it is interesting to note that systems of the form $R$FeWO${_6}$ have recently been reported to be multiferroic \cite{Somnath}. However, these $R$FeWO${_6}$ systems crystallize in the polar $Pna2{_1}$ structure, and exhibit low temperature multiferroic transitions between 15-18K, which involve a complex interplay between the Fe and the magnetic $R$ ions. 
Fig.\ref{Fig5}(a) depicts the magnetic field dependence of the real part of the dielectric susceptibility $\epsilon'(T)$ as measured in $\gamma$-Fe${_2}$WO${_6}$ at 0 and 5 Tesla magnetic fields. A suppression of $\epsilon'(T)$ is observed on the application of the magnetic field. This data is replotted in the form of the magneto-dielectric effect ($|{\Delta \epsilon}/\epsilon{_0}| \%$) in 
\ref{Fig5}(b), and shows that a finite magneto dielectric effect is observed right up to room temperatures. Though the extent of magneto-dielectricity as measured at 10 kHz is of the order of 5 $\%$, the fact that $|{\Delta \epsilon}/\epsilon{_0}|$ decreases with the probing frequency suggests that the contribution of the magnetoresistive component could be appreciable. The inset shows the variation of the same quantity in the region of the low temperature glass transition., showing that the low temperature glass like transition is also seen to be associated with a discontinuity in the magneto-dielectric effect, in-spite of the fact that the values of $|{\Delta \epsilon}/\epsilon{_0}|$ are very small. 

\begin{figure}
	\includegraphics[scale=0.405]{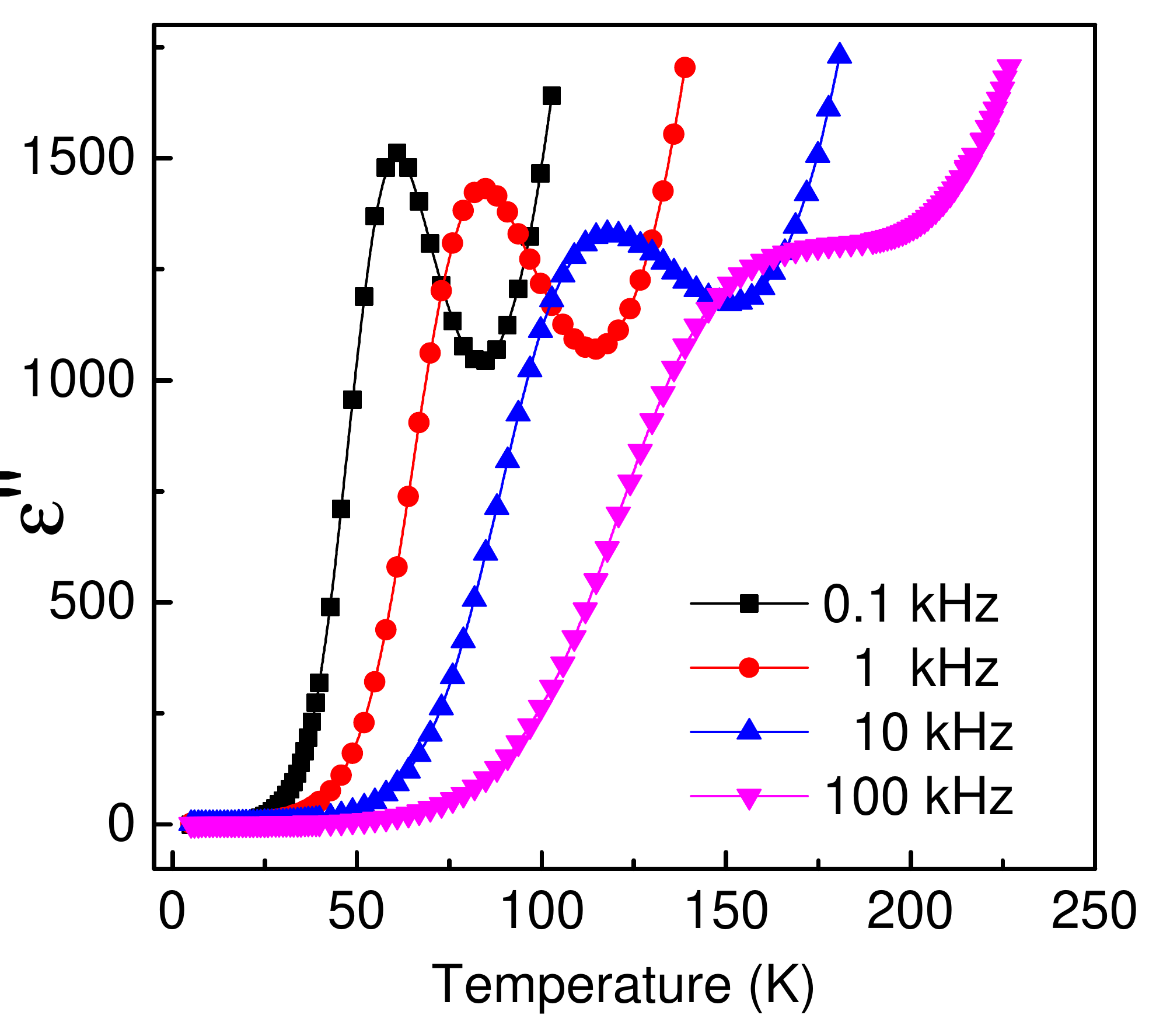}
	\caption{Temperature dependence of the dielectric loss $\epsilon''(T)$ at different probing frequencies as measured in $\gamma$-Fe${_2}$WO${_6}$. A frequency dependent poeak, typical of charge relaxation processes is observed.}
	\label{Fig6}
\end{figure}

The temperature dependence of the imaginary part of the dielectric susceptibility exhibits a pronounced frequency dependent peak, the peak position of which shifts to higher temperatures with increase in the probing frequencies as shown in Fig.(\ref{Fig6}). These is typical of a charge relaxation process, and similar signatures have been observed in a number of other other strongly correlated oxides, especially within the magnetic state of the phase separated manganites \cite{Freitas}. In the mixed valent manganites, electronic phase separation typically gives rise to regions with ferromagnetic and antiferromagnetic correlations, the competition between which is known to result in glassiness in both the magnetic and electronic sectors. This electronic phase separation which results in phase co-existence over a wide range of length scales then manifests itself in the form of frequency dependent dielectric properties. In the case of the  $\gamma$-Fe${_2}$WO${_6}$ system investigated here, this dispersion observed in the dielectric susceptibility could arise as a consequence of the oxygen non-stoichoimetry. This could also be intimately tied with the low temperature magnetic glass state which sets in at lower temperatures. For instance, early Electron Paramagnetic Resonance (EPR) measurements have speculated on the possibility of magnetic clusters (presumably arising fro anti-site disorder) in the Fe${_2}$WO${_6}$ polymorphs\cite{GuskosJSSC} . More recent measurements \cite{Guskos} have also indicated the presence of short range correlations extending well above the magnetic ordering temperatures in the $\beta$-Fe${_2}$WO${_6}$ polymorph, indicating that this cluster formation could be generic to the Fe-W-O systems. 

In summary, we report on the synthesis and characterization of the $\gamma$-Fe${_2}$WO${_6}$ system. Crystallizing in the non-polar $Pbcn$ space group, this system exhibits a number of magnetic transitions as a function of decreasing temperatures, with long range antiferromagnetic order setting in at temperatures as high as 280K. These transitions, especially the high temperature ones, are also observed to be associated with a finite magneto-dielectric effect. The presence of a low temperature magnetic glass like state is also inferred from aging and magnetic viscosity measurements. These presumably arise as a consequence of the freezing of magnetic clusters, the existence of which is also suggested by the frequency dependent peak in the imaginary part of the measured dielectric susceptibility.   
\section{Acknowledgments}
The authors thank D. Buddhikot for help in heat capacity measurements. J.K. acknowledges DST India for support through a SERB-NPDF file no. PDF/2016/000911 . S.N. acknowledges DST India for support through grant no. SB/S2/CMP-048/2013.

\bibliography{Bibliography}

\end{document}